\documentclass[%
twocolumn,
amsmath,amssymb,
aps, physrev,
]{revtex4-2}

\usepackage{graphicx}
\usepackage{dcolumn}
\usepackage{bm}
\usepackage{xcolor}

\begin{document}

\preprint{APS/123-QED}

\title{\textbf{The title should simply and concisely convey the main findings. Avoid nonstandard abbreviations and acronyms.} 
}%

\title{Phase-Resolved Ultra-High-Energy Emission Defines an Energetic Boundary\\
for Compact-Object Engines and Dense-Matter Structure}

\author{Marlon M. S. Mendes}
\email{marlon.mendes.102043@ga.ita.br}
\affiliation{Departamento de F\'isica e Laborat\'orio de Computa\c{c}\~ao Cient\'ifica Avan\c{c}ada e Modelamento (Lab-CCAM), Instituto Tecnol\'ogico de Aeron\'autica, DCTA, 12228-900, S\~ao Jos\'e dos Campos, SP, Brazil}

\author{C\'esar H. Lenzi}
\affiliation{Departamento de F\'isica e Laborat\'orio de Computa\c{c}\~ao Cient\'ifica Avan\c{c}ada e Modelamento (Lab-CCAM), Instituto Tecnol\'ogico de Aeron\'autica, DCTA, 12228-900, S\~ao Jos\'e dos Campos, SP, Brazil}

\date{\today}


\begin{abstract}

We show that phase-resolved ultra-high-energy emission from
LS I \(+61^\circ303\)
defines a critical energetic boundary linking the compact-object
engine, circumstellar interaction depth, and dense-matter structure.
Combining the GeV--TeV--UHE spectral transition with an orbital
acceleration--transparency sieve and branch-resolved stellar sequences
for 116 certified EoS, we find that the measured \(25\)--\(100\) TeV component remains
energetically viable for all 116 canonical-mass EoS and therefore does
not discriminate among them. EoS discrimination arises only in the
deliberately more demanding pure-hadronic \(>1\) TeV stress-test
construction:
 at 2.0 kpc and \(1.4\,M_\odot\), this boundary gives
\(I_{45,\rm crit}=1.5504\), with 82 of 116 EoS satisfying the
energetic condition at\(\dot P_{\rm ref}=3\times10^{-15}\,{\rm s\,s^{-1}}\).
A separately applied tidal diagnostic, \(\Lambda_{1.4}\leq580\), is satisfied by 43 EoS, with nine satisfying both conditions. These models span several microscopic classes but
occupy a narrow structural region, indicating structural overlap rather
than selection of a unique EoS family. A separate GeV power-budget test requires
\(\dot P_{\rm rot}\gtrsim
5.65\times10^{-14}f_\Omega\,{\rm s\,s^{-1}}\)
at 2.0 kpc. For \(f_\Omega\sim1\), the source lies well above the UHE
crossing and the EoS discrimination disappears. The UHE observation
therefore identifies the engine--environment regime in which
differences in neutron-star structure can become observationally
discriminating, rather than defining a universal dense-matter selector.

\end{abstract}

\maketitle


\section{Introduction}

The equation of state (EoS) of cold supranuclear matter remains a
central problem in relativistic astrophysics. Massive pulsars require
viable EoS models to support neutron stars near or above
\(2\,M_\odot\)
\cite{Demorest2010,Antoniadis2013,Cromartie2019,Fonseca2021};
NICER constrains the mass--radius relation
\cite{Riley2019,Miller2019,Riley2021,Miller2021}; and GW170817
constrains the tidal response
\cite{Abbott2017,Abbott2018}. Yet substantially different high-density
compositions remain compatible with current multimessenger data.

We ask whether this residual structural diversity can be exposed by an
independently constructed high-energy energetic constraint. We use a
heterogeneous library assembled primarily from CompOSE
\cite{Typel2015,Oertel2017,CompOSE2022} and external tables. After
source, support, stability, and branch-resolution checks, 116 EoS are
retained, spanning nucleonic, relativistic mean-field, crossover, and
hybrid descriptions, without prior weighting by microscopic family.

LS I \(+61^\circ303\) provides such a laboratory. It exhibits
orbitally modulated GeV emission from \textit{Fermi}-LAT
\cite{FermiLSI}, TeV emission from MAGIC and VERITAS
\cite{MAGICLSI,VERITASLSI}, and UHE emission reported by LHAASO
\cite{LHAASOLSI}.
FAST also detected transient radio pulsations at
\(P=269.15508\pm0.00016~{\rm ms}\), providing direct evidence for a
rotating neutron star \cite{Weng2022}. At UHE energies, acceleration,
photon escape, hadronic interaction efficiency, and compact-engine
power must be satisfied simultaneously.

In this Letter these requirements are kept causally separate. Orbital
acceleration and \(\gamma\gamma\)-transparency are independent of the
dense-matter EoS; dense matter enters the energetic boundary only
through the moment of inertia \(I(M)\) in the rotation-powered
spin-down luminosity. Radius, tidal deformability, and other TOV--Love
observables are retained only as downstream diagnostics. Because the
intrinsic phase-connected \(\dot P_{\rm rot}\) is unknown, we invert the
UHE requirement into critical boundaries in spin-down, hadronic
interaction depth, and \(I(M)\). At fixed source conditions,
\(\dot P_{\rm crit}\propto I(M)^{-1}\), so EoS differences become
discriminating only near energetic saturation.

The measured \(25\)--\(100\) TeV mixed component is primarily
environmental and remains viable throughout the certified
canonical-mass cohort under the reference conditions. A more demanding
pure-hadronic \(>1\) TeV construction instead crosses the physical
\(I(M)\) distribution at 2.0 kpc. Its intersection with a separately applied tidal diagnostic identifies a narrow structural region spanning several
microscopic classes, rather than a unique EoS family. If the
phase-averaged GeV and UHE components must share the same approximately
isotropic rotation-powered reservoir, however, the required intrinsic
spin-down lies above the UHE crossing and the EoS discrimination
disappears. Thus the UHE observation identifies an
engine--environment regime in which neutron-star structure can become
observationally discriminating, rather than a universal dense-matter
selector.


\section{UHE Energetic Boundary}
\label{sec:energetic_boundary}

Broadband inputs are taken from \textit{Fermi}-LAT,
MAGIC/VERITAS, and LHAASO
\cite{FermiLSI,MAGICLSI,VERITASLSI,LHAASOLSI}. For each orbital
realization we require
\begin{equation}
\frac{E_{\max}}{E_{\rm req}}>1,
\qquad
\tau_{\gamma\gamma}(100~{\rm TeV})<1,
\end{equation}
and define \(f_{\rm joint}\) as the fraction satisfying both
conditions.

This acceleration--transparency sieve is evaluated independently of
the dense-matter EoS.

Dense matter enters only through the stellar moment of inertia,
\begin{equation}
{\rm EoS}\rightarrow I(M)\rightarrow
\dot E_{\rm sd}
=
4\pi^2 I(M)\frac{\dot P_{\rm rot}}{P^3}.
\label{eq:engine_chain}
\end{equation}
We use 116 certified EoS with native support over the target-mass
range and branch-resolved TOV--Love--Hartle sequences. No stellar
quantity is extrapolated beyond native tabular support; radius and
tidal observables are retained only as downstream diagnostics.

FAST measured
\(P=269.15508\pm0.00016~{\rm ms}\)
\cite{Weng2022}. Its apparent folding derivative is not a
phase-connected intrinsic spin-down
\cite{Weng2022,Suvorov2022}. We therefore use
\(\dot P_{\rm ref}=3\times10^{-15}~{\rm s\,s^{-1}}\)
\cite{LHAASOLSI} only as an energetic benchmark and express the
physical result as a boundary in the unknown intrinsic
\(\dot P_{\rm rot}\).

For a pulsar-powered hadronic channel, the UHE luminosity depends on
the spin-down power, proton loading, and hadronic interaction efficiency. Requiring the proton-loading
efficiency not to exceed unity defines
\begin{equation}
\dot P_{{\rm crit},i}(M)
=
\frac{K_{\rm src}}{I_i(M)},
\label{eq:inverseI_main}
\end{equation}
where \(K_{\rm src}\) contains only source and environmental
quantities. Thus, at fixed source conditions, the energetic ordering
is exactly the ordering by decreasing \(I(M)\). Coherent source
systematics can shift the boundary but cannot reorder the EoS; this
rank is therefore an energetic ordering, not a posterior preference
score.

We consider two energetic requirements. The measured
\(25\)--\(100\) TeV mixed component corresponds to
\(L_{\gamma,25-100}=5.03\times10^{31}~{\rm erg\,s^{-1}}\)
at 2.0 kpc and primarily probes the circumstellar interaction depth.
A more demanding pure-hadronic \(>1\) TeV construction gives
\(L_{\gamma,>1{\rm TeV}}=1.672\times10^{33}~{\rm erg\,s^{-1}}\).
At the adopted disk-density ceiling and
\(\dot P_{\rm rot}=\dot P_{\rm ref}\), the latter corresponds to
\(I_{45,\rm crit}=1.5504\) at \(1.4\,M_\odot\).

We evaluate distances of 2.0 kpc and
\(2.65\pm0.09\) kpc
\cite{LHAASOLSI,Lindegren2021,Weng2022},
and masses \(M=1.4,\ 1.8,\) and \(2.0\,M_\odot\).
Spectral-normalization and index uncertainties are propagated with
\(10^7\) Monte Carlo draws, while instrumental calibration
uncertainties are treated as seven coherent bounded states. The
resulting calibration bands are therefore systematic envelopes rather
than probability intervals.

Independent mass--radius and tidal constraints
\cite{Fonseca2021,Riley2019,Riley2021,Abbott2018}
are applied only after the energetic calculation. In particular,
\(\Lambda_{1.4}\leq580\) does not modify
\(f_{\rm joint}\), the hadronic interaction efficiency, or
\(\dot P_{\rm crit}\); its intersection with the UHE boundary is used
only as a downstream structural diagnostic.

The full expressions for the hadronic luminosity, proton-loading
threshold, \(pp\) interaction efficiency, critical moment of inertia,
and the mixed and pure-hadronic flux constructions are given in the
Supplemental Material ~\cite{Supplemental}.


\section{Results}

The results follow three physical layers: spectral decomposition,
orbital viability, and the compact-engine energetic boundary. Dense
matter enters only in the last step through the EoS-dependent moment
of inertia \(I(M)\).

\begin{figure*}[t]
    \centering
    \includegraphics[width=0.80\textwidth]{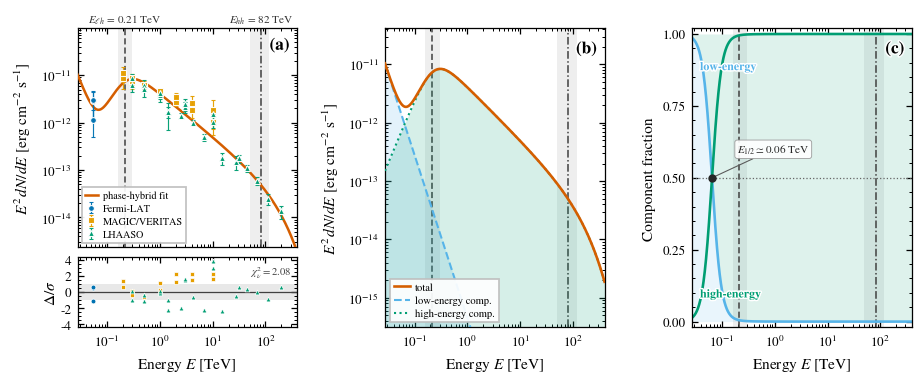}
    \caption{
    Broadband spectral transition of LS I \(+61^\circ303\).
    (a) Multi-instrument spectral energy distribution and phase-hybrid
    fit, with \(E_{\ell h}\simeq0.21~{\rm TeV}\) and
    \(E_{hh}\simeq82~{\rm TeV}\).
    (b) Low- and high-energy components.
    (c) Relative contributions of the two components.
    }
    \label{fig:sed_transition}
\end{figure*}

The broadband transition is shown in
Fig.~\ref{fig:sed_transition}. The phase-hybrid decomposition separates
the GeV--sub-TeV component from that sustaining the TeV--UHE tail,
with characteristic scales
\(E_{\ell h}\simeq0.21~{\rm TeV}\) and
\(E_{hh}\simeq82~{\rm TeV}\).
The latter defines the energetic requirement studied below.

\begin{figure*}[t]
    \centering
    \includegraphics[width=0.90\textwidth]{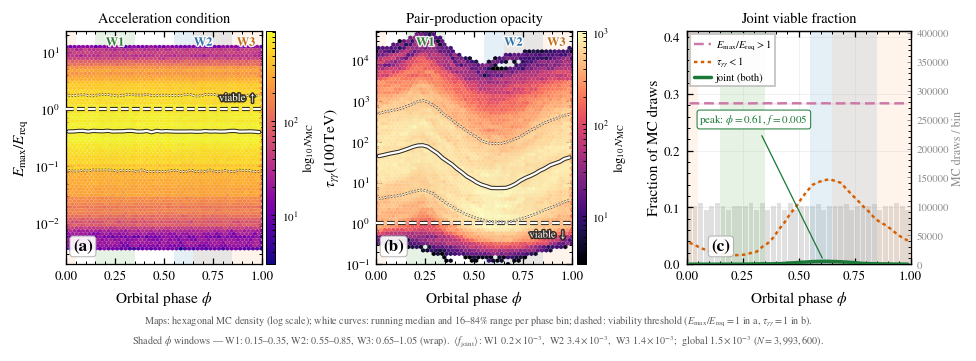}
    \caption{
    Phase-resolved UHE orbital viability.
    (a) Acceleration condition \(E_{\max}/E_{\rm req}\).
    (b) Pair-production opacity
    \(\tau_{\gamma\gamma}(100~{\rm TeV})\).
    (c) Fractions satisfying acceleration, transparency, and both;
    the overlap peaks near \(\phi\simeq0.61\).
    }
    \label{fig:orbital_mc}
\end{figure*}

The orbital calculation restricts where the UHE component can be
produced and escape. As shown in Fig.~\ref{fig:orbital_mc},
the acceleration and transparency conditions overlap only in a narrow
phase interval, with
\(f_{\rm joint}^{\rm max}\simeq5\times10^{-3}\) near
\(\phi\simeq0.61\), compared with
\(\langle f_{\rm joint}\rangle\simeq1.5\times10^{-3}\).
This phase-localized window is obtained independently of the
dense-matter EoS.

\begin{figure*}[t]
    \centering
    \includegraphics[width=0.98\textwidth]
    {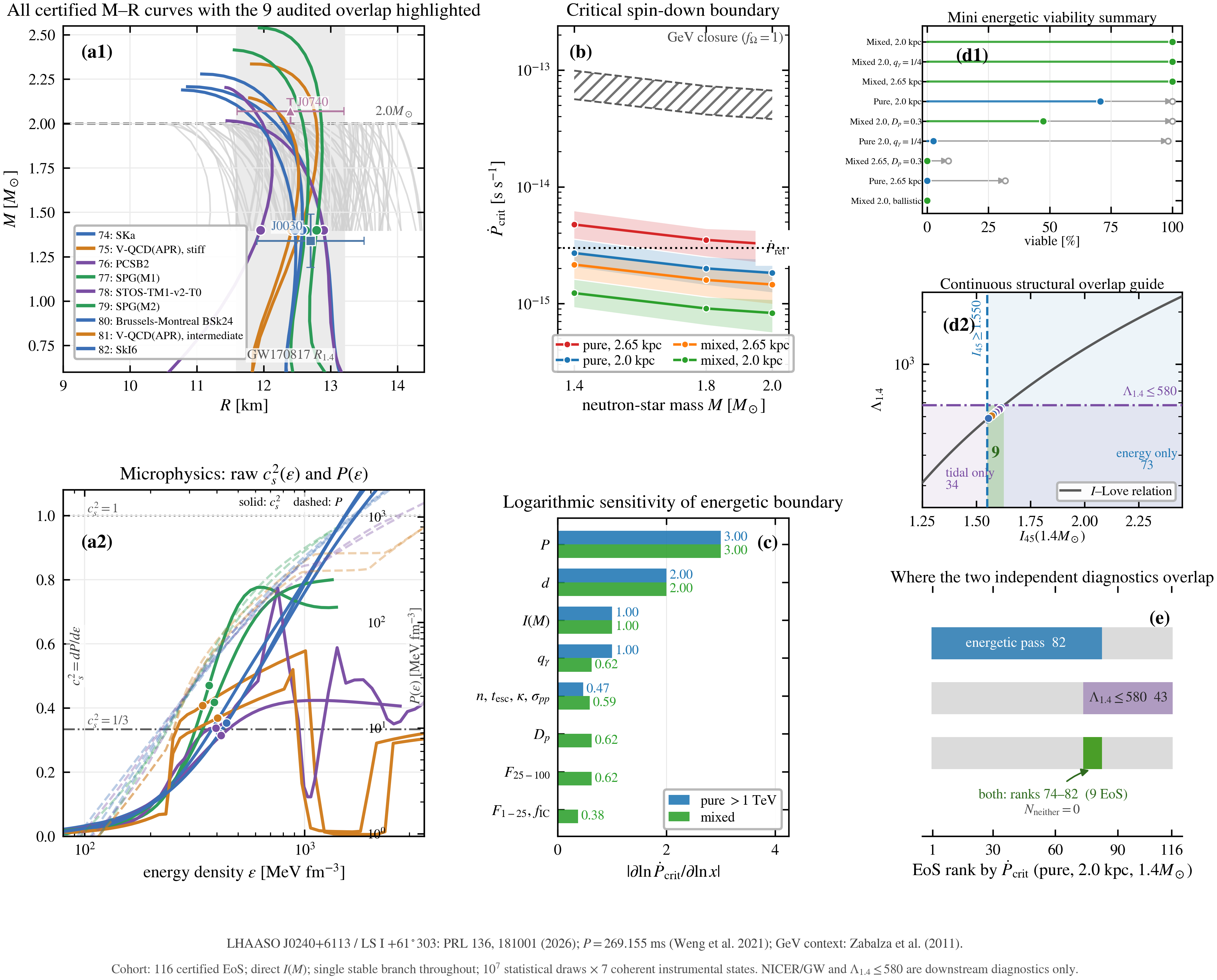}
    \caption{
    Energetic boundary and downstream structural diagnostics for the
    116 certified EoS.
    (a1) Mass--radius sequences with the nine energetic--tidal overlap
    models highlighted.
    (a2) Corresponding \(c_s^2(\varepsilon)\) and \(P(\varepsilon)\).
    (b) Critical spin-down boundaries versus mass; solid curves show
    the cohort medians and shaded bands the EoS spread at the central
    spectral parameters. Statistical \(N_0,\Gamma\) uncertainties and
    bounded instrumental systematics are propagated separately
    (Supplemental Material).
    (c) Logarithmic sensitivities of the energetic boundary.
    (d1) Energetic viability across the certified cohort.
    (d2) Separately applied energetic and tidal thresholds in the
    \(I_{45}\)--\(\Lambda_{1.4}\) plane.
    (e) Rank-space intersection for the pure-hadronic \(>1\) TeV
    stress test at 2.0 kpc and \(1.4\,M_\odot\): 82 EoS satisfy the
    energetic condition, 43 satisfy \(\Lambda_{1.4}\leq580\), and
    nine satisfy both.}
\label{fig:uhe_boundary_structure}
\end{figure*}

The compact-engine and structural results are summarized in
Fig.~\ref{fig:uhe_boundary_structure}. For the measured
\(25\)--\(100\) TeV mixed tail,
\(\dot P_{\rm ref}=3\times10^{-15}~{\rm s\,s^{-1}}\)
requires median-EoS minimum densities of
\(2.45\times10^{10}\) and
\(4.35\times10^{10}~{\rm cm^{-3}}\) at 2.0 and 2.65 kpc,
respectively. These values lie below the adopted disk density but
above the stellar-wind density.

Under these reference conditions, the measured
\(25\)--\(100\) TeV component remains energetically viable for all
116 certified canonical-mass EoS and therefore does not discriminate
among them. EoS discrimination arises only in the deliberately more
demanding pure-hadronic \(>1\) TeV stress test. At 2.0 kpc,
\(M=1.4\,M_\odot\), and the disk-density ceiling, this construction
crosses the physical \(I(M)\) distribution at
\(I_{45,\rm crit}=1.5504\), with 82/116 EoS satisfying the energetic
condition; at 2.65 kpc none do. These counts refer to the tested
library and are not EoS probabilities.

The shaded bands in Fig.~\ref{fig:uhe_boundary_structure}(b) show the
EoS spread in \(\dot P_{\rm crit}\) at the central spectral
parameters, not statistical or instrumental uncertainty bands.
The statistical \(N_0,\Gamma\) errors are propagated separately with
\(10^7\) spectral realizations: for the pure-hadronic \(>1\) TeV
case, the 16th--84th percentile propagation rescales
\(\dot P_{\rm crit}\) by \(0.891\)--\(1.115\) relative to the
central spectrum. Instrumental systematics are treated separately as
coherent bounded scenarios; details and numerical intervals are given
in the Supplemental Material.

The crossing shifts with mass because \(I(M)\) increases along the
stellar sequences. At fixed mass,
\(\dot P_{{\rm crit},i}\propto I_i^{-1}\), so coherent source
systematics move the energetic boundary without changing the EoS
ordering. The sensitivity analysis in
Fig.~\ref{fig:uhe_boundary_structure}(c) identifies the pulsar period
and distance as the leading source-side levers, while \(I(M)\)
contains the direct dense-matter dependence.

A separately applied tidal diagnostic provides a complementary structural
test. Of the 116 certified EoS, 43 satisfy
\(\Lambda_{1.4}\leq580\), and nine satisfy both this condition and
the central pure-hadronic energetic boundary. These nine occupy ranks
74--82 and cluster within
\(I_{45}(1.4)=1.555\)--\(1.614\),
\(R_{1.4}=11.95\)--\(12.90~{\rm km}\), and
\(\Lambda_{1.4}=489\)--\(563\).
The intersection therefore identifies a narrow structural overlap
spanning several microscopic classes, rather than a uniquely preferred
EoS family.

The interpretation remains engine dependent. The phase-averaged GeV
luminosity requires
\(\dot P_{\rm rot}\gtrsim
5.65\times10^{-14}f_\Omega~{\rm s\,s^{-1}}\)
at 2.0 kpc for the median moment of inertia. Thus, if the GeV and UHE
components share the same rotation-powered reservoir with
\(f_\Omega\sim1\), the intrinsic spin-down lies well above the UHE
crossing and the EoS discrimination disappears.

The UHE observation therefore does not define a universal EoS
selector. It identifies the restricted engine--environment regime in
which differences in \(I(M)\) become observationally discriminating.

\section{Conclusion}

We have shown that phase-resolved UHE emission from
LS I \(+61^\circ303\) defines an energetic boundary connecting the
compact-object engine, circumstellar interaction depth, and
dense-matter structure. The orbital acceleration--transparency
constraint is independent of the EoS, while dense matter enters the
energetic problem only through the stellar moment of inertia \(I(M)\).
Consequently, EoS differences become discriminating only when the
system lies sufficiently close to energetic saturation.

For the measured \(25\)--\(100\) TeV mixed tail, the required
interaction depth is compatible with the adopted disk environment and
the energetic condition remains viable throughout the certified
canonical-mass cohort under the reference assumptions.

Thus, the measured \(25\)--\(100\) TeV component does not discriminate
among the 116 certified canonical-mass EoS. The \(82/116\)
EoS-discriminating result arises only from the deliberately more
demanding pure-hadronic \(>1\) TeV stress-test construction:

at 2.0 kpc and \(1.4\,M_\odot\),
\(I_{45,\rm crit}=1.5504\), with 82 of 116 certified EoS satisfying
the corresponding energetic condition.

A separately applied tidal diagnostic,
\(\Lambda_{1.4}\leq580\), is satisfied by 43 EoS, with nine satisfying
both conditions. These models span several microscopic classes but
occupy a narrow structural region, demonstrating a structural overlap
rather than selection of a unique EoS family.

The interpretation remains explicitly engine dependent. Direct GeV
closure requires
\(\dot P_{\rm rot}\gtrsim
5.65\times10^{-14}f_\Omega~{\rm s\,s^{-1}}\)
at 2.0 kpc. If the GeV and UHE components share the same
approximately isotropic rotation-powered reservoir
(\(f_\Omega\sim1\)), the source lies well above the UHE crossing and
the EoS discrimination disappears. The UHE observation therefore does
not define a universal dense-matter selector; it identifies the
engine--environment regime in which differences in \(I(M)\) can become
observationally discriminating.

A phase-connected measurement of the intrinsic spin evolution of
LS I \(+61^\circ303\), together with improved constraints on its mass
and circumstellar environment, would determine whether the system
actually occupies this regime.
\section*{Acknowledgments}
This study was financed in part by the Coordenação de Aperfeiçoamento de Pessoal de N\'ivel Superior – Brasil (CAPES) – Finance Code 001. This work has been done as part of the Project INCT-F\'isica Nuclear e Aplica\c{c}\~oes, under No. 408419/2024-5. It was also supported by Conselho Nacional de Desenvolvimento Cient\'ifico e Tecnol\'ogico (CNPq) under Grants No. 305327/2023-2 (C.H.L.), No. 307255/2023-9 (O.L.), No. 01565/2023-8~(Universal - O.L, C.H.L), No. 409736/2025-2~(Universal - O.L, C.H.L.), No. 444797/2024-6 (O.L.), and Funda\c{c}\~ao de Amparo \`a Pesquisa do Estado de S\~ao Paulo (FAPESP) under Thematic Project No. 2024/17816-8 (O.L., C.H.L.).

\bibliography{apssamp}
\end{document}